\documentclass[%
 reprint,
superscriptaddress,
 amsmath,amssymb,
 aps,
]{revtex4-2}

\usepackage[english]{babel}
\usepackage[utf8x]{inputenc}
\usepackage[T1]{fontenc}

\usepackage{graphicx}
\usepackage{dcolumn}
\usepackage{bm}
\usepackage{multirow}
\usepackage{xcolor}

\usepackage{xcolor}

\begin{document}

\preprint{APS/123-QED}

\title{Vulnerabilities of quantum key distribution systems in visible range}
\author{Boris Nasedkin}
  \email{banasedkin@itmo.ru}
  \affiliation{Laboratory of Quantum Processes and Measurements, ITMO University, 199034, 3b Kadetskaya Line, Saint Petersburg, Russia}
  \affiliation{Laboratory for Quantum Communications, ITMO University, 199034, 3b Kadetskaya Line, Saint Petersburg, Russia}
    
 \author{Azat Ismagilov}%

\affiliation{Laboratory of Quantum Processes and Measurements, ITMO University, 199034, 3b Kadetskaya Line, Saint Petersburg, Russia}
 
  \author{Vladimir Chistiakov}

\affiliation{Laboratory for Quantum Communications, ITMO University, 199034, 3b Kadetskaya Line,
  Saint Petersburg, Russia} 

\author{Andrei Gaidash}%

\affiliation{Laboratory of Quantum Processes and Measurements, ITMO University, 199034, 3b
  Kadetskaya Line, Saint Petersburg, Russia}

\author{Aleksandr~Shimko}%

\affiliation{Center for Optical and Laser materials research, Saint-Petersburg State University,
  198504, 5 Ulianovskaya, Saint Petersburg, Russia}

 \author{Alexei D. Kiselev}%

\affiliation{Laboratory of Quantum Processes and Measurements, ITMO University, 199034, 3b
  Kadetskaya Line, Saint Petersburg, Russia} 
  \affiliation{Laboratory for Quantum Communications, ITMO University, 199034, 3b Kadetskaya Line, Saint Petersburg, Russia}

\author{Anton Tcypkin}

\affiliation{Laboratory of Quantum Processes and Measurements, ITMO University, 199034, 3b
  Kadetskaya Line, Saint Petersburg, Russia}

\author{Vladimir Egorov}

\affiliation{Laboratory for Quantum Communications, ITMO University, 199034, 3b Kadetskaya Line,
  Saint Petersburg, Russia}

 \author{Anton Kozubov}

\affiliation{Laboratory of Quantum Processes and Measurements, ITMO University, 199034, 3b
  Kadetskaya Line, Saint Petersburg, Russia}

\date{\today}

\begin{abstract}
  In this paper we investigate spectral vulnerabilities in quantum key distribution systems arising from the use of shorter-wavelength radiation in the 400–800 nm range, with particular focus on the induced photorefraction attack (IPA). Crucial elements influenced by IPA include various types of modulators, both phase and
intensity modulators. In the following paper, we consider different scenarios and their
implications. Through combined theoretical and experimental analysis, we demonstrate that optical components commonly used as countermeasures in the telecom band (1000–2100 nm) exhibit significantly reduced effectiveness at shorter wavelengths. The efficiency of IPA is shown to increase as the wavelength decreases, posing a substantial threat to phase-modulation-based QKD protocols. We analyze the impact of IPA across different QKD architectures and assess the feasibility of potential countermeasures under realistic implementation scenarios. Our results highlight the necessity of broadband security evaluations and wavelength-aware component design in future QKD systems.
\end{abstract}

\maketitle

\section{Introduction}\label{sec1}

\label{sec:introduction}

The security of quantum key distribution (QKD) systems hinges upon the fundamental laws of physics. Over the past decades, significant progress has been made in theoretical security
proofs~\cite{Scarani:rmp:2009,Gisin:rmp:2002,renner2005information,portmann2014cryptographic,shor2000simple,christandl2004generic,inamori2007unconditional,ben2005universal,leverrier2013security,curty2009upper,gottesman2004security,lo2005decoy}. However, many of these proofs rest on the assumption of perfect and trusted devices. Such trust assumption imposes a rather strong security requirement, which is generally difficult to implement in practice. In real situations, imperfect devices in QKD systems would introduce vulnerabilities that can be exploited by eavesdroppers to extract additional information about the transmitted bits during their attacks, as it is considered in \cite{gottesman2004security,lo2005decoy,gaidash2022subcarrier,fung2008security,zhang2021security,bochkov2019security}. Device-independent (DI) \cite{vazirani2019fully,zapatero2023advances,arnon2018practical,arnon2019simple,chistiakov2019feasibility,lu2022unbalanced} and semi-device-independent \cite{lo2012measurement,braunstein2012side,lucamarini2015practical,pirandola2017fundamental} protocols have emerged as a prominent area of research that provides a way to prevent side-channel attacks~\cite{Acin:prl:2007}. On the opposite side, another rapidly developing field which is often referred to as quantum hacking has received a considerable amount of attention~\cite{lu2023hacking}. The field of quantum hacking focuses on identifying and addressing vulnerabilities in the technical implementation of QKD systems.
The significance of researching hardware loopholes in QKD systems arises from their transition from laboratory prototypes to commercial deployment. As QKD technologies advance toward practical implementation, identifying and mitigating potential hardware vulnerabilities becomes crucial to ensuring the security and reliability of real-world quantum communication networks~\cite{gaidash2022quantum,zhou2022quantum,illiano2022quantum,ghalaii2020capacity}.

Emerging approaches in quantum hacking pose novel challenges for security analysis of QKD systems. Numerous vulnerabilities have been identified in prior works \cite{sajeed2021approach,sun2022review,makarov2024preparing}, underscoring the importance of enhancing security measures \cite{lu2023experimental,lu2021intensity,alferov2023studying,zhang2021securing}. Eavesdropper strategies in quantum hacking can broadly be classified under two basic types, active \cite{lovic2023quantified,lydersen2011controlling,qian2018hacking} and passive \cite{meda2017quantifying,gisin2006trojan}, as well as combinations thereof. Active strategies involve the manipulation of optical parameters within the hardware, whereas passive strategies solely facilitate the acquisition of additional information. 

A quintessential illustration of a passive attack is the Trojan-horse attack (THA), which is also known as the large pulse attack~\cite{vakhitov2001large, gisin2006trojan}. In this scenario, the eavesdropper (Eve) employs scanning pulses introduced into the systems of legitimate users (Alice and Bob) via the quantum channel, allowing for the extraction of distributed sequences from the reflected portions of the eavesdropper pulses back into the quantum channel. The key factors that determine the efficiency of this attack involve the transmittance of elements in two directions, the distinguishability of states along with
countermeasures utilized in the system. An important point is that all these factors require the eavesdropper to define the optimal spectral range of wavelengths employed to perform attack.

The well-known detector blinding attack (DBA)~\cite{makarov2009controlling,chistiakov2019controlling, gao2022ability} exemplifies an active attack. The idea underlying the attack is that, under the action of high-power laser radiation (as compared to typical intensities used in QKD), the operating mode of single photon detectors (SPDs) based on avalanche photodiodes can be changed from the Geiger mode into the linear operating mode, so that the eavesdropper can manipulate the clicks registered by the SPDs. The efficiency of this DBA based quantum hacking approach crucially depends on the spectral sensitivity of the avalanche photodiode, the transmittance of optical elements connecting the quantum channel with the single photon avalanche diode (SPAD), and the countermeasures implemented into the QKD system. So, it turned out that, similar to the Trojan-horse attack, the spectral range is one of the factors that govern the success of the DBA attack. 

Previous studies~\cite{borisova2020risk,nasedkin2023loopholes,sushchev2021practical,jain2014risk, sushchev2024trojan} motivated by the THA, have mainly been focused on transmittance of fiber optical elements utilized in QKD systems at wavelengths ranged between 1000~nm and 2100~nm. For the 600-1000 nm range, the experimental results were reported in Refs.~\cite{jain2014risk,nasedkin2022analyzing}.

But, for the above discussed DBA and the laser damage attacks (LDA) that exploit temperature dependence of the transmittance of fiber-optical elements~\cite{alferov2022study,huang2020laser,ponosova2022protecting}, shorter wavelengths may also be used. At shorter wavelengths, low insertion losses of elements employed for countermeasures  may lead to unnoticed and thus potentially successful delivery of radiation injected by Eve to the elements under attack.

New vulnerabilities at different wavelengths have been the subject of intense recent studies~\cite{chaiwongkhot2022faking}. The induced-photorefraction attack (IPA)~\cite{lu2023hacking,han2023effect} represent attacks that will do better in the visible spectral range~\cite{ye2023induced}. In addition, this attack is claimed to be applicable for hacking measurement device independent (MDI) QKD systems~\cite{lu2023hacking}.

The main targets of the IPA are phase and amplitude modulators based on photo-refractive crystals, where variations in refractive index or transmittance can be induced by light. The IPA strategy enables Eve to manipulate the characteristics of quantum states, i.e. mean photon number, at the sender's side using laser radiation. The efficiency of these attacks improves as the radiation wavelength becomes shorter~\cite{ye2023induced}. So, it is important to perform  spectral measurements of fiber components used in QKD systems at shorter wavelengths so as to identify and mitigate vulnerabilities exposed by the attack. In particular, the 400-800 nm spectral range has not been received a proper attention yet. In this paper, our goal is to fill the gap.

Note that, according to Ref.~\cite{ye2023induced}, successful implementation of the IPA requires low intensities starting from 3~nW. A comparison between these intensities and the maximal intensity permissible for insertion into a fiber without causing damage, which is typically around 9~W (this value is applicable to quartz single mode fibers without power damage level doping commonly used in fiber communications and may vary among manufacturers) leads to the conclusion that attenuation needed for passive defence against IPA for single pass through optical elements can roughly be estimated at about 100~dB. Thus, in order to assess the feasibility of the attack, it is necessary to measure and analyze insertion losses of individual optical elements on the side of Alice. To this end, we study the insertion losses of different optical elements in the 400--800 nm spectral range, demonstrate the existence of potential vulnerabilities and suggest potential countermeasures.

The paper is organized as follows.
Security issues arose due to considered effect are discussed in Sec.~\ref{subsec:estimation}.
In Sec.~\ref{sec:experimental-setup} we describe our experimental technique employed to measure spectra of insertion losses of optical elements at wavelengths ranged from 400~nm to~800~nm. 
The experimental data for insertion losses measured for a variety of optical attenuators, wave-division multiplexing (WDM) and isolating components are presented in Sec.~\ref{sec:experimental-results}.
Sec.~\ref{sec:ipa} is dedicated to countermeasures discussion.
Finally, Sec.~\ref{sec:conclusion} concludes the article.

\section{Security issues}
\label{subsec:estimation}

In considerations of side-channel attacks aimed at extracting quantum information, the visible spectral range is often disregarded on account of  its poor confinement in single-mode fibers. Nevertheless, this range plays a significant role in the manipulation of optical components.
Techniques such as laser damage and induced photorefraction exploit visible-wavelength radiation to impair or alter the functionality of critical devices without directly accessing the quantum channel. 

In this work, we focus on the induced-photorefraction attack (IPA) as a representative example of such vulnerabilities. We assess its potential impact across a range of QKD protocols and system architectures described in the literature. Particular focus is placed on phase and intensity modulators, whose disruption poses a direct threat to state preparation fidelity. 
This section highlights these vulnerabilities and outline their broader consequences for QKD system security.

Change of phase modulator's (PM's) modulation index may be considered separately, depending on implemented optical devices.
A PM can be a part of the following commonly used in the
field of QKD devices: 
\begin{enumerate}
    \item PMs introducing a constant phase shift within Mach–Zehnder interferometer (MZI) schemes; 
    \item PMs with an oscillating phase shift, as utilized in subcarrier wave (SCW) QKD
      \cite{gaidash2022subcarrier,sajeed2021approach,miroshnichenko2018security,chistiakov2019feasibility}; 
    \item PMs integrated into variable optical attenuators (VOAs);
    \item  PMs employed as key elements in amplitude modulators (AMs) or intensity modulators (IMs).
\end{enumerate}

In addition to the above classification, QKD protocol and utilized quantum states play essential role in security estimations.
Accordingly, we distinguish the following categories of QKD schemes:
\begin{enumerate}
    \item Protocols utilizing linearly dependent states (such as single-photon BB84
      \cite{bennett2014BB84} or single-photon fraction in decoy-state protocols
      \cite{lo2005decoy,ma2005practical,wang2005beating}), where possible underestimation of key parameters or phase-remapping attack \cite{fung2007phase} should be taken into consideration.
The security problem for such protocols has already been studied in a number of papers\cite{ye2023induced,han2023effect,lu2023hacking}; 
    \item Protocols with linearly independent set of states (coherent states with phase-coding \cite{bennett1992quantum,lo2006security,bennett1992experimental,gobby2004quantum} or SCW QKD
      have not been considered so far in the light of IPA auxiliary attack.
      In what follows, we concentrate on these kinds of protocols.
\end{enumerate}

\subsection{Linearly independent set of states}
Now the second group of the protocols listed above will be our primary concern. We begin with the protocols utilizing coherent states with phase-coding\cite{bennett1992quantum,lo2006security,bennett1992experimental,gobby2004quantum}.


Roughly speaking, when a nonlinear material such as lithium niobate is illuminated with the pump light, the photorefractive effect gives rise to the light-induced change of the refractive index arising through the linear electro-optic (Pockels) effect from the space-charge electric field caused by inhomogeneous redistribution of photogenerated charges
(see, e.g., \cite{lu2023hacking}; and references therein for more details). The photorefractively changed refractive index leads to the photoinduced shift of the signal phase that, in turn, affects electro-optic response and the working states of lithium niobate-based devices widely employed in QKD systems.

In particular, owing to this phase shift, the voltage modulation curves of MZI-based VOA and IM appear to be displaced along the voltage axis introducing deviations of the DC-bias voltage. This effect has been previously employed in the context of IPAs against the measurement-device-independent (MDI) QKD \cite{ye2023induced} the decoy-state BB84 QKD \cite{lu2023hacking} and the Gaussian-modulated coherent-state continuous-variable QKD in \cite{zhou2025security}.

In our case, however, the assumption that the only photorefraction-induced effect is the phase shift identical for all states does not provide any advantage to an eavesdropper. The reason is that all the states are equally distributed in the phase plane and such phase shift does not affect the phase difference between the states.

From the other hand, one of the experimentally measured results reported in \cite{han2023effect} is the photorefractively induced change of the half-wave voltage in PMs that can be described as the photoinduced change of the modulation index. According to Ref. \cite{han2023effect}, this effect is similar to the phase-remapping attack \cite{fung2007phase}, causing the phase difference between coded states to change.

In this case, phase-remapping by a factor of $x$ of is applicable: 
\begin{gather}
   \{ |\alpha_0\rangle,\ |\alpha_0 e^{\frac{i\pi}{N}}\rangle,\ |\alpha_0
   e^{\frac{i2\pi}{N}}\rangle,\ ...\ ,\ |\alpha_0 e^{\frac{i(2N-1)\pi}{N}}\rangle\}
   \rightarrow\notag\\  \rightarrow \{|\alpha\rangle,\ |\alpha e^{\frac{ix\pi}{N}}\rangle,\ |\alpha
   e^{\frac{i2x\pi}{N}}\rangle,\ ...\ ,\ |\alpha
   e^{\frac{i(2N-1)x\pi}{N}}\rangle\},\label{phaseremap} 
\end{gather}
where $\alpha_0$ is the complex amplitude of the coherent state, $\alpha$ is the complex amplitude after
IPA with $|\alpha|\le|\alpha_0|$, $2N$ is the total number of states in the set.
After the IPA, the eavesdropper can apply additional commonly used attack,
e.g. unambiguous state discrimination (USD)
attack.

This attack is characterized by the USD probability, $P_{U}(\alpha,x)$, 
giving the optimal probability for USD of a set of states.
When the states are equiprobable, this probability can be
evaluated by computing the minimal eigenvalue of
the Gram matrix, $G$~\cite{Sun:pra:2002,Horoshko:pla:2019,kozubov2021quantum,gaidash2022subcarrier}.
For the unperturbed coherent states,
which are symmetrically arranged, 
the eigenvalues can be obtained in the closed form
using the discrete Fourier transform technique
(see, e.g.~\cite{kozubov2021quantum}).
After the IPA, the Gram matrix,
which is a positive semidefinite Toeplitz matrix,
is no longer represented by a circulant matrix and
computing the eigenvalues requires numerical analysis.

\begin{figure*}[ht]
\centering
\includegraphics[width=0.9\textwidth]{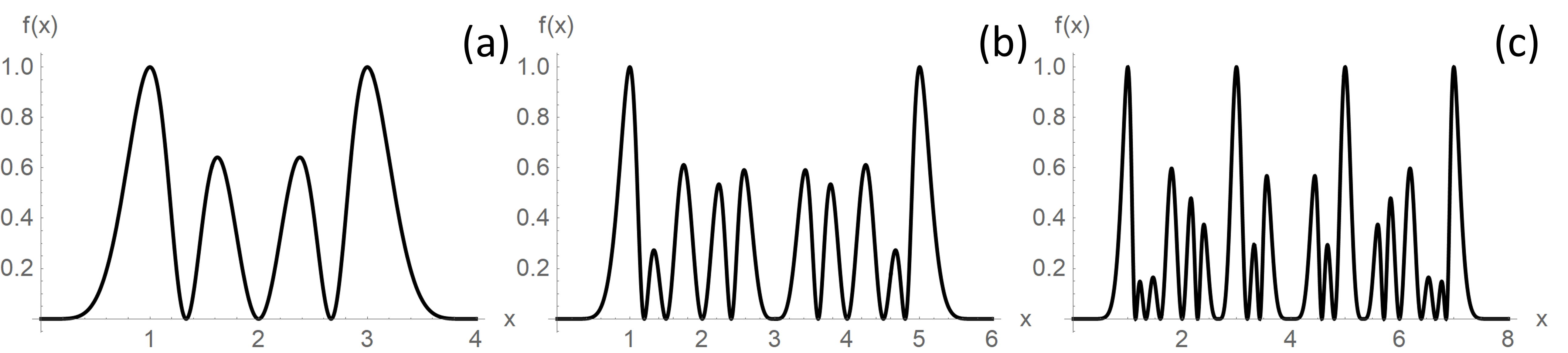}
\caption{
 USD probability ratio $f(x)=P_{U}(\alpha,x)/P_{U}(\alpha,1)$
  computed as a function of the phase-remapping parameter $x$
  (see Eq.~\eqref{phaseremap}) at (a)~$N=2$, (b)~$N=3$, (c)~$N=4$.
  The curves show that phase-remapping with non-integer $x$
  leads to
  a significant reduction in unambiguous state discrimination probability.
 }  
\label{fig:fx}
\end{figure*}

However, a few analytical observations can be made.
In particular, we have found that
the ratio of the probabilities $f(x)=P_{U}(\alpha,x)/P_{U}(\alpha,1)$
does not exceed the unity, $0\le f(x)\le 1$.
In addition, for $N>1$, it weakly depends on $|\alpha|$.
For the special case
with just two states as in B92 and
$N=1$, we have 
\begin{gather}
    P_{U}(\alpha,x)=1-e^{-|\alpha|^2(1-\cos(\pi x))}
\end{gather}
and
\begin{gather}
    f(x)=\frac{1-e^{-|\alpha|^2(1-\cos(\pi x))}}{1-e^{-2|\alpha|^2}}.
  \end{gather}
The results for
the USD probability ratio computed
as a function of $x$
at $N=2$, $N=3$ and $N=4$
are shown in Fig.~\ref{fig:fx}.
Also note, that 
\begin{gather}
    P_U(\alpha,1)\approx \frac{2N(|\alpha|^2)^{2N-1}}{(2N-1)!}.
\end{gather}
The overlaps of the states after the IPA only increase, so it cannot provide any significant
advantage to an eavesdropper.

\subsection{Intensity variation in SCW QKD}
\begin{figure}[h]
\centering
\includegraphics[width=0.45\textwidth]{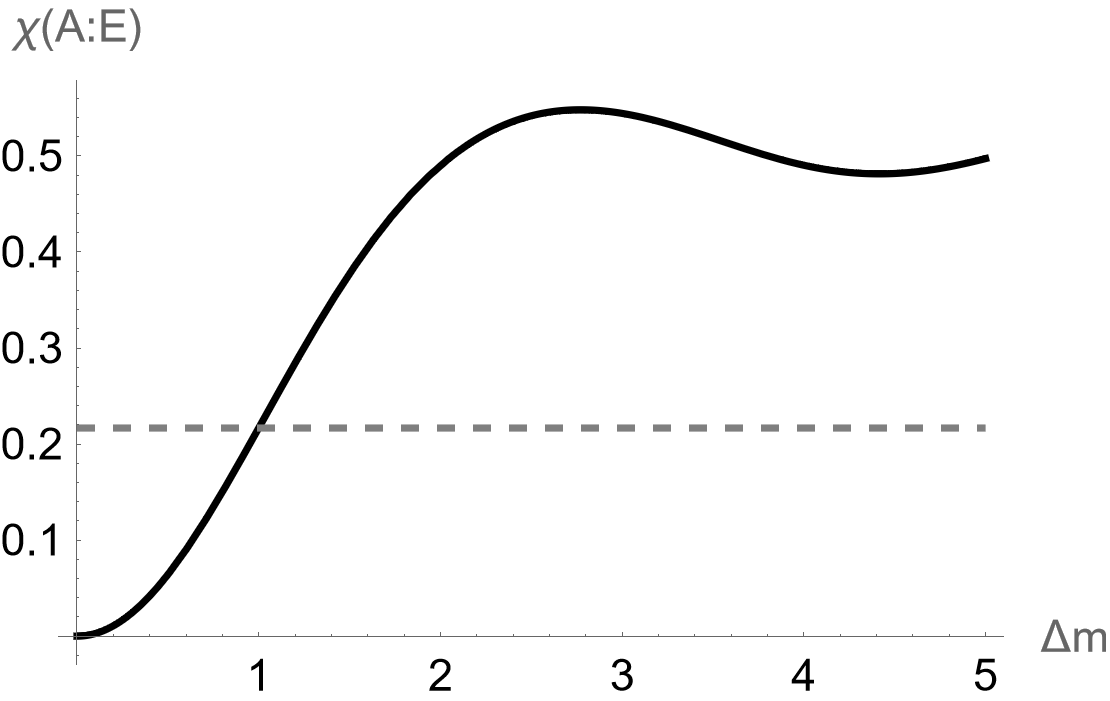}
\caption{Variation of Holevo bound $\chi(A:E)$ as in Eq.~\eqref{holevo} depending on the (multiplicative) factor $\Delta m$ of modulation index $m$, denoted as solid black line, compare to the initial value of the Holevo bound with $\Delta m=1$, denoted as dashed gray line; calculations were performed for the following values: $|\alpha_0|^2=1$, $m=0.434$ that determines the ratio between the power of the sidebands to the power of the carrier to be equal $1:10$
 }  
\label{fig:scwholevo}
\end{figure}

As for SCW QKD setup, IPA here plays the same role as attacking an IM or a VOA based on a PM in
protocols with coherent states and phase-coding, since total power of sidebands is a
  function of modulation index $m$ , more precisely, 
 \begin{gather}
     |\alpha|^2=|\alpha_0|^2( 1-J_0^2(m))\approx \frac{|\alpha_0|^2m^2}{2},
 \end{gather}
where approximation is valid for small values of $m<1$ and $J_0(m)$ is Bessel function of the first
kind), and $|\alpha_0|^2$ is the total power. And the modulation index can be increased by utilizing the IPA with a recovery, as described in
\cite{han2023effect}, where legitimate parties recalibrate their setups after IPA. 
In turn, an increase of the modulation index significantly enhances the information accessible to an eavesdropper, as it can be seen in Fig.~\ref{fig:scwholevo}. This quantity is known as Holevo bound, and it can be formally evaluated as follows \cite{miroshnichenko2018security}:

\begin{gather}
    \chi(A:E)=h\Big(\frac{1-e^{-|\alpha_0|^2( 1-J_0^2(2m))}}{2}\Big).\label{holevo}
\end{gather}

\subsection{Intensity variation in decoy-state}

 A similar concern emerges in the implementation of decoy-state QKD protocols \cite{lo2005decoy}, where precise parameter estimation—particularly of the total gain $Q_\mu$, the photon-number-dependent yield $Y_n$, single-photon error fraction $E_1$ and total error $E_{\mu}$ —is vital for security proofs. These parameters depend on the mean photon number $\mu$, for which tight bounds are required. However, most security frameworks assume that $\mu$ remains stable and immune to external manipulations. However, under the influence of induced photorefraction, this assumption may no longer hold. The external manipulation of optical intensity by an adversary can lead to an artificial increase in $\mu$, resulting in erroneous parameter estimation and potential security compromise, as discussed in~\cite{han2023effect, ye2023induced}.

Additionally, such intensity modifications open the door to phase-remapping attacks \cite{fung2007phase}. Despite proposed approaches  were reported to be suboptimal, they are feasible with current technological level.

These considerations highlight the protocol- and implementation-specific nature of IPA vulnerabilities. As a result, a comprehensive security analysis must include accurate optical characterizations across both infrared and visible spectral regimes.





\section{Experimental approach}
\label{sec:experimental-setup}

While the previous section addressed the general security implications of IPA, it is essential to recognize that such effects manifest only under well-defined optical conditions. A realistic assessment of IPA threats in QKD systems requires quantifying the probe light power that can reach susceptible components, particularly LiNbO3-based modulators.
In general, assuming that the maximal power of radiation that can propagate in silica optical fibers without damage is about 40 dBm \cite{kashyap2013fuse,davis1997comparative,huang2020laser}, the above power of probing radiation can be evaluated using the following expression:

\begin{equation}
  P_{\mathrm{dBm}}(\lambda) = 40-\sum_{i=1}^n (\alpha_{\mathrm{n}}(\lambda)),
\label{eq:pow}
\end{equation}
where $P_{\mathrm{dBm}}(\lambda)$ is the estimated power of Eve’s source which will reach the modulator under the attack, an ($\lambda$) are insertion losses of the
nth fiber-optical element (the elements are ordered along the path of
Eve’s radiation).

Clearly, using this formula requires a systematic characterization of the optical components’ response in the visible spectral range.
This section presents an experimental framework for evaluating these characteristics and examines the corresponding implications for system design and countermeasure implementation.

\begin{figure}[h]
\centering
\includegraphics[width=0.45\textwidth]{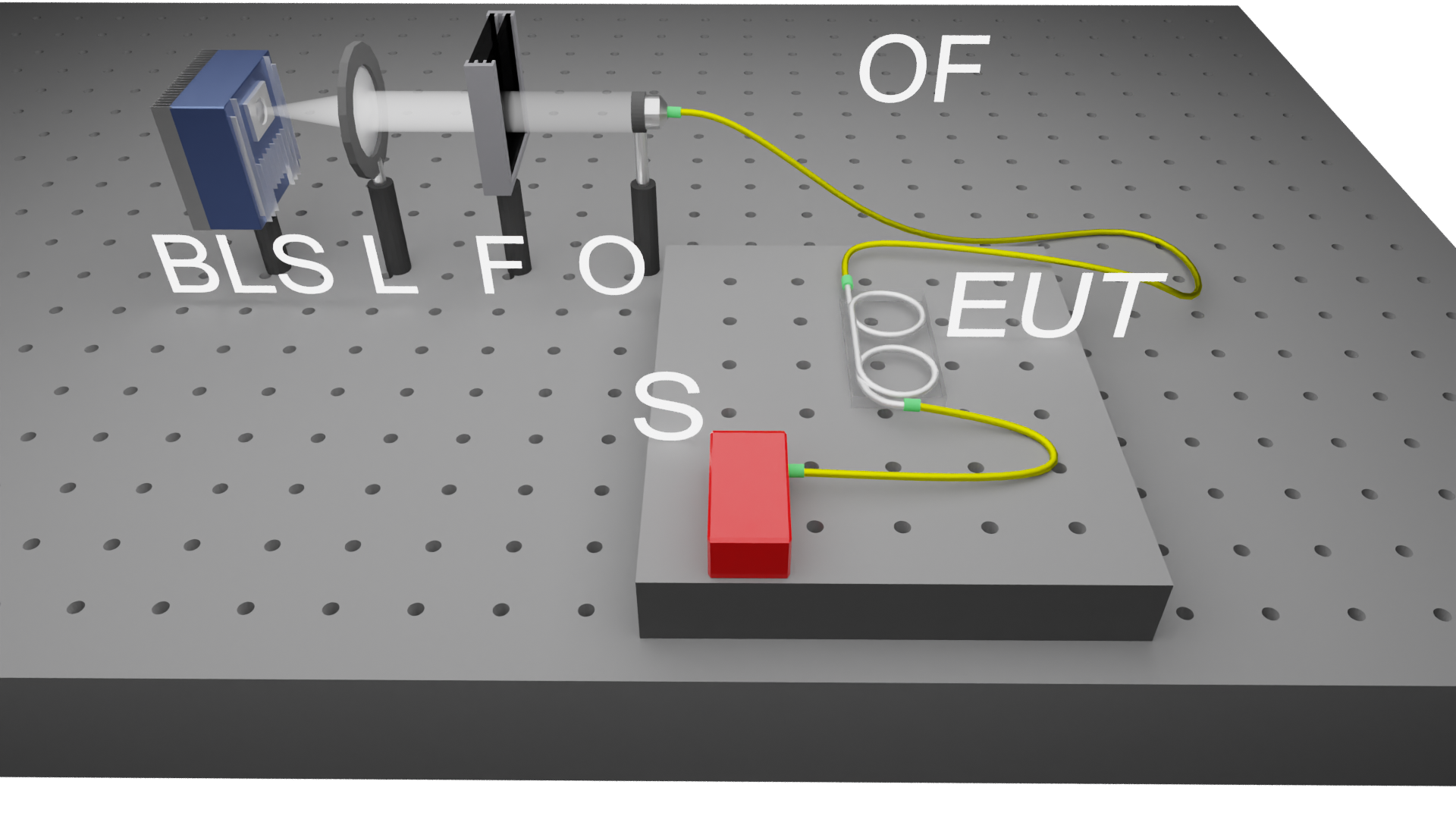}
\caption{Optical scheme for insertion losses measurements in the visible range. BLS is broadband light source, L is lens, F is set of filters, O is objective, OF is single mode optical fiber, EUT is element under test, S is spectrometer}
\label{scheme}
\end{figure}

Figure~\ref{scheme} shows the optical scheme of our experimental setup used to measure the insertion losses of optical elements in the visible range. In this scheme, a broadband plasma light source (XWS65, ISTEQ)
indicated as the broadband light source (BLS)  is utilized to produce light  with wavelengths ranged from 250~nm up to 2500~nm. Referring to Fig.~\ref{scheme}, the radiation from this source is directed into a single-mode fiber ITU-T G.652 (OF) passing through a set of lenses (L), filters (F) and the objective (O) which is positioned on a three-axis stage to meet coupling and spectrum requirements. After the fiber, the focused light propagates through either a fiber or a fiber with the optical elements under test and  is measured by the spectrometer (USB4000-UV-VIS-ES, Ocean Optics) with the 1.5~nm resolution.

An important point is that the spectrometer is equipped with its own multimode fiber with an SMA905 connector. So, in order to align the single-mode and multimode fibers, we have used the FC-SMA905 connector.

Due to variations in efficiency of coupling to the single-mode fiber, losses due to absorption by the filters and technical restrictions imposed by the spectrometer, the measured spectrum of BLS was narrowed to the 400-800 nm range. Insertion losses of the optical element under test were calculated as follows: 
\begin{equation}
  \alpha_{\mathrm{dB}}(\lambda) = -10
  \log_{10}\bigl[
  P_{\mathrm{mes}}(\lambda)/(P_{\mathrm{ref}}(\lambda)T_{\mathrm{f}}(\lambda))
  \bigr],
\label{eq:T_dB}
\end{equation}
where the powers $P_{\mathrm{ref}}(\lambda)$ and$P_{\mathrm{mes}}(\lambda)$ are  measured at wavelength $\lambda$ without and with the tested optical element, respectively; $T_{\mathrm{f}}(\lambda)$ is the transmission of the set of filters. Note that, in our experiments, the neutral filters were used to extend the dynamic range of measurements. For each element, we have performed ten measurements and the experimental results were averaged (details on the error estimation procedure can be found in Appendix~B of Ref.~\cite{nasedkin2023loopholes}). In what follows, we concentrate on the optical elements that are usually placed on the sender's side between the output of the phase and amplitude modulators and the entrance of the quantum channel. 

The optical scheme behind our experimental technique enables us to perform measurements of the
insertion losses with approximately  $50$~dB dynamic range. It is simple to use and low-cost when
compared with analogous schemes discussed, e.g., in Ref.~\cite{makarov2024preparing}. An important
point is that, by using suitably modified methods presented in Ref.~\cite{sushchev2021practical},
the proposed scheme can be improved so as to extend both the wavelength range and the dynamic range
of measurements. As far as certification of QKD systems is concerned, this is a feasible and
easy-to-use scheme that can be employed for characterization of optical elements' insertion losses.

\section{Experimental results}
\label{sec:experimental-results}

\subsection{Optical attenuators}
\label{subsec:attenuators}

In this section, we present the results for two variable optical attenuators (VOAs) that differ in the mode of operation, along with six different fixed optical attenuators (FOAs).

In QKD systems, optical attenuators (OA) are typically employed to reduce the intensity of laser pulses to the single photon level. Additionally, OAs are of importance in reducing the probability of wavelength-dependent attacks successful realization. Consequently, there is a requirement to avoid pronounced variations of spectral insertion loss properties of OAs across a broad wavelengths range.

In QKD systems, the insertion losses of VOAs are typically adjusted by applying voltage. This implies that during the key distribution process, the attenuation value can be tuned in the range  between the maximum and minimum insertion losses. To address this variability, we have evaluated the maximal and minimal insertion losses settings for one electro-mechanical (EM) and one electro-optical (EO) model of VOA. According to the manufacturer datasheets, at the wavelength $\lambda=1550$~nm,  the minimal values of insertion losses without power (the field free regime without applied voltage) are 0.77~dB and 0.80~dB for EO VOA and EM VOA, respectively. When the applied voltage is 5~V, the corresponding measured values of insertion losses for EO VOA and EM VOA are 36.0~dB and 43.4~dB.

\begin{figure*}[ht]
\centering
\includegraphics[width=0.9\textwidth]{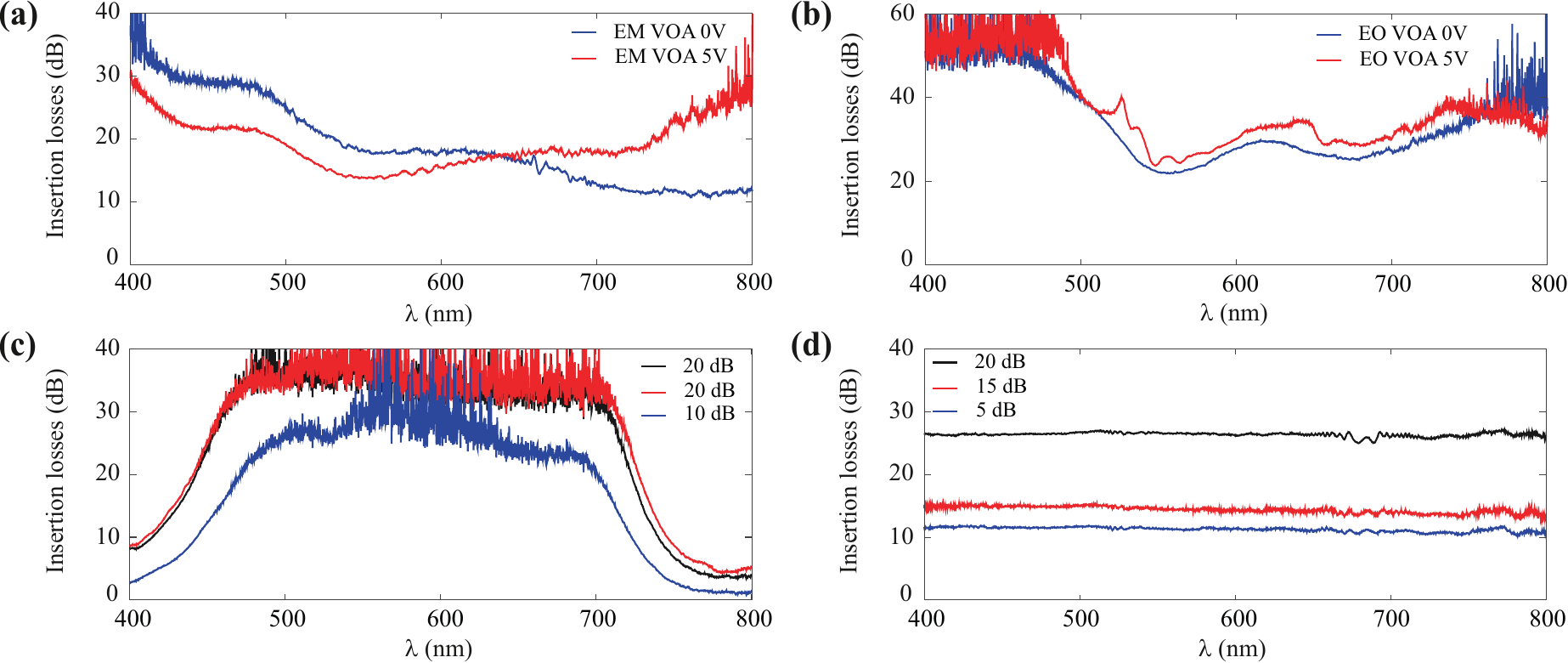}
\caption{Insertion losses measured in the 400-800 nm range for (a)~electro-mechanical and (b)~electro-optical  variable optical attenuators (blue and red solid lines indicate the zero-voltage curve and the case where the applied voltage is 5~V, respectively); (c)~absorption based fixed attenuators (the curves for the first 20~dB FOA, second 20~dB FOA and 10~dB FOA are indicated by black, red, and blue solid lines, respectively); (d)~scattering based fixed attenuators principles (the curves for 20~dB FOA, 15~dB FOA and 5~dB FOA are indicated by black, red, and blue solid lines, respectively).
} 
\label{fig:Att}
\end{figure*}

Figures~\ref{fig:Att}(a) and~(b) present wavelength dependencies of insertion losses measured in the 400-800 nm range for EM VOA and EO VOA at two values of applied voltage: 0~V and 5~V.

In the zero-voltage case, it turned out that the measured insertion losses are higher than the values reported in the datasheet. For EM VOA and EO VOA, the minimal losses 10.7~dB and 21.8~dB are reached at the wavelengths 761~nm and  557~nm, respectively.

Referring to Figs.~\ref{fig:Att}(a) and~(b), when the applied voltage is 5~V (blue lines), the losses are shown to vary in the range between 13.7~dB and 30~dB for EM VOA, whereas, for EO VOA, they are ranged from
23.7~dB to 50~dB.  Note that these maximum values of losses are approximate, as they correspond to the noise level of our experimental setup and depend on the number of neutral filters.

Insertion losses measured  in relation to the wavelength for six FOAs are shown in Figs.~\ref{fig:Att}(c) and~(d). For three absorption based FOAs (see Figs.~\ref{fig:Att}(c)), according to the values reported by the manufacturers, the insertion losses of one FOA is 10~dB, whereas the losses for the other two FOAs are the same and equal to  20~dB. For the scattering based FOAs (see Figs.~\ref{fig:Att}(d)), there are three different reported values of losses at the operating wavelength ($\lambda=1550$~nm): 5~dB, 15~dB and 20~dB. These values are in good agreement with the results of our measurements performed at $\lambda=1550$~nm. 


The insertion losses of the absorption based FOAs measured in the 400-800 nm range are shown in Fig.~\ref{fig:Att}(c). It can be seen that, for each FOA, there are two transmittance windows where the insertion losses are noticeably lower than the value at the operating wavelength indicated in the legend. The window ranged between 400~nm and 430~nm is located in the vicinity of local minima for the two 20~dB attenuators with $\alpha_{\mathrm{dB}}\approx 8.1$ and $\alpha_{\mathrm{dB}}\approx 8.6$ at the wavelength $405$~nm (for the 10~dB attenuator, the corresponding value is $\alpha_{\mathrm{dB}}\approx 2.7$) Similarly, the other window in the 750-800~nm wavelengths range is close to the local minima at $\lambda\approx 780$~nm with the insertion losses estimated at $\alpha_{\mathrm{dB}}\approx 3.4$ and $\alpha_{\mathrm{dB}}\approx 4.3$ and $\alpha_{\mathrm{dB}}\approx 0.9$.

Figure~\ref{fig:Att}(d) shows the wavelength dependencies of the insertion losses measured for the scattering based FOAs. It is seen that, for the 20 dB attenuator, the insertion losses insignificantly change with wavelength, varying from 27~dB to 25~dB. A similar conclusion applies to the 15~dB (5~dB) attenuator, where variations are between 15.6~dB (11.2~dB) and 13 dB (10.8~dB).

\subsection{Wavelength-division multiplexing components}
\label{subsec:wdm}

\begin{figure*}[ht]
\centering
\includegraphics[width=0.9\textwidth]{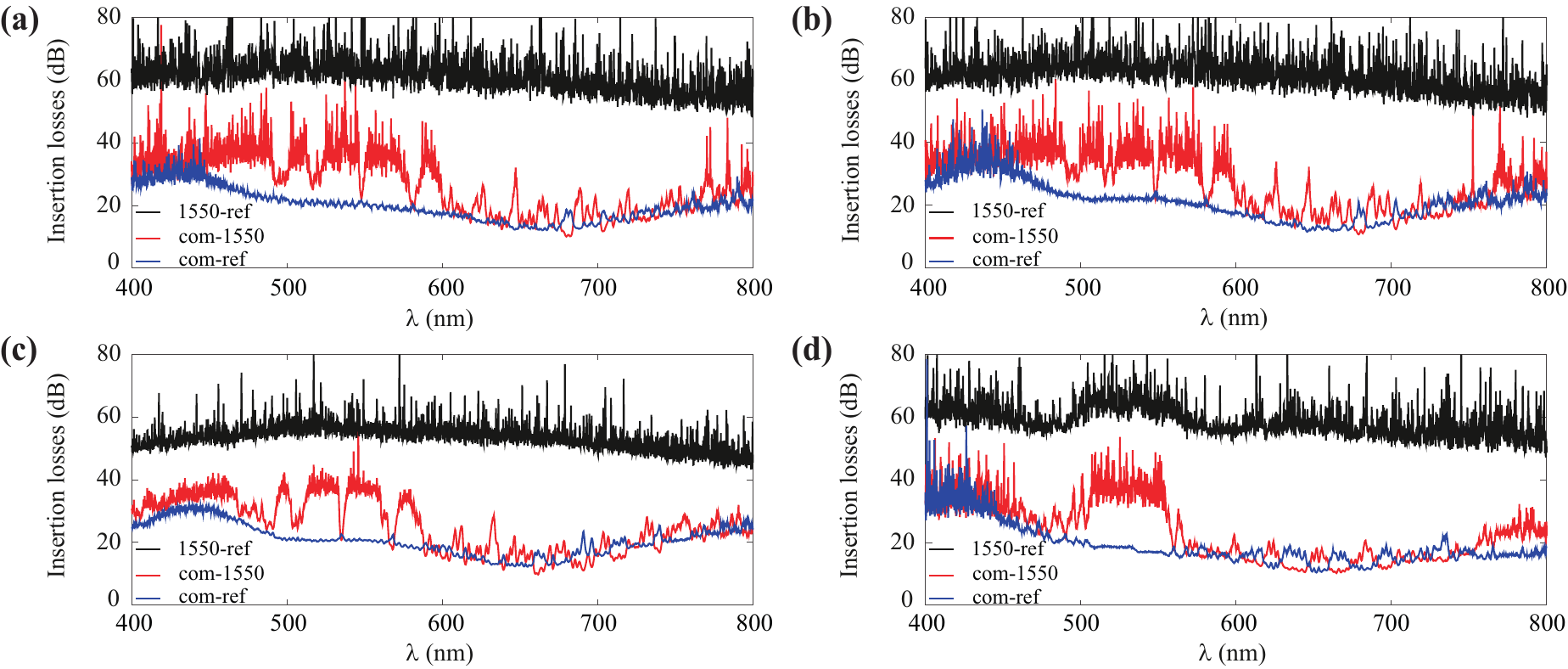}
\caption{Insertion losses measured in the 400-800 nm range for (a)~CWDM1; (b)~CWDM2; (c)~CWDM3 and (d)~DWDM (black, blue, and red solid lines indicate insertion losses measured for light propagating along three different directions: from the ref-port to the 1550-port, from the com-port to the ref-port and  from the com-port to the 1550-port, respectively)} 
\label{fig:CWDM}
\end{figure*}

In QKD systems, wavelength-division multiplexing (WDM) components can be used as countermeasures by suppressing radiation with wavelengths outside the range utilized by the communication channel between
Alice and Bob. In the simplest case, the key elements of the internal design of a fiber WDM component are thin films deposited on a substrate. Ideally, at the working wavelength, light  will enter the common port (the com-port) and, after passing through the thin films and the substrate,  will propagate to the corresponding port of the operating wavelengths grid. Other wavelengths are directed to the reflecting port (the ref-port). This port might be connected with a watchdog detector that monitor Eve's probing pulses. The latter assumes that the sensitivity range of the watchdog detector is broad enough to cover wavelengths of potential attacks.

In practice, there are a number of factors limiting the efficiency  of the WDM-components. These include the substrate transmittance as well as the number, ordering, and the thicknesses of the thin films. Differences between the perfectly designed mode of operation and its practical implementation may lead to side channels available to the eavesdropper. In our previous study~\cite{nasedkin2023loopholes}, the WDM-components are found to have the transmittance windows outside the telecom range, in agreement with the theoretical calculations reported
in~\cite{gu2006design}.

Similar to Ref.~\cite{nasedkin2023loopholes}, we consider both coarse WDM (CWDM) and dense WDM (DWDM) components. Specifically, insertion losses in the 400-800~nm range are measured for the four filters: three CWDM-components from the same manufacturer where two of them belong to the same production batch (CWDM1 and CWDM2) and one DWDM-component. The results are shown in Fig.~\ref{fig:CWDM}.

In addition to the com-port and the ref-port, the WDM components have the 1550-port corresponding to the operating wavelength $\lambda=1550$~nm. Referring to Figure~\ref{fig:CWDM}, when the probing light propagates from the com-port to the 1550-port, each WDM-component  reveals the ranges of wavelengths where the insertion losses are within the interval between $10$~dB and $20$~dB. Minimal values of insertion losses are presented in Table \ref{tab:WDM}. Note that, for the WDM-components under consideration, the losses in the telecom range at non-working wavelengths are about $30$~dB.

\begin{table}[htp]
    \centering
    \begin{tabular}{|p{2.5cm}|p{2.5cm}|p{2.5cm}|}
      \hline
      Component &  $\alpha_{\mathrm{min}}$ (dB) & $\lambda_{\mathrm{min}}$ (nm)\\
      \hline
    CWDM1 &  10.0 & 680\\
    CWDM2 & 10.7 & 679\\
    CWDM3 &  9.7 & 662\\
      DWDM & 10.3 & 665\\
      \hline
    \end{tabular}
    \caption{Minimal insertion losses and corresponding wavelengths for WDM components.}
    \label{tab:WDM}
\end{table}

From Fig.~\ref{fig:CWDM}, it can be seen that, for the light propagating from the com-port to the ref-port, the insertion losses vary between 10~dB and 30~dB which is close to the values for the above discussed direction from the com-port to the 1550-port. For the case of light propagation from the 1550-port to the ref-port, it turned out that the losses are close to the noise level. Note that,  when directions of light propagation are reversed, changes in insertion losses appear to be well below accuracy of measurements (the results of these measurements are omitted) and thus nonreciprocity effects have not been observed in our experiments. 

\subsection{Isolating components}
\label{subsec:isolating}

Isolating elements are typically utilized to reduce reflections back into laser resonators. Similar to the WDM components, in QKD systems, these elements may  serve as countermeasures against attacks.

In this section, we examine two elements: the dual stage isolator and the circulator. Since we are mainly interested in countermeasures against the IPA attack,  the case of backward propagation of injected light where  probing pulses travel from the output port to the input port will be our primary concern.   

\begin{figure*}[ht]
\centering
\includegraphics[width=0.9\textwidth]{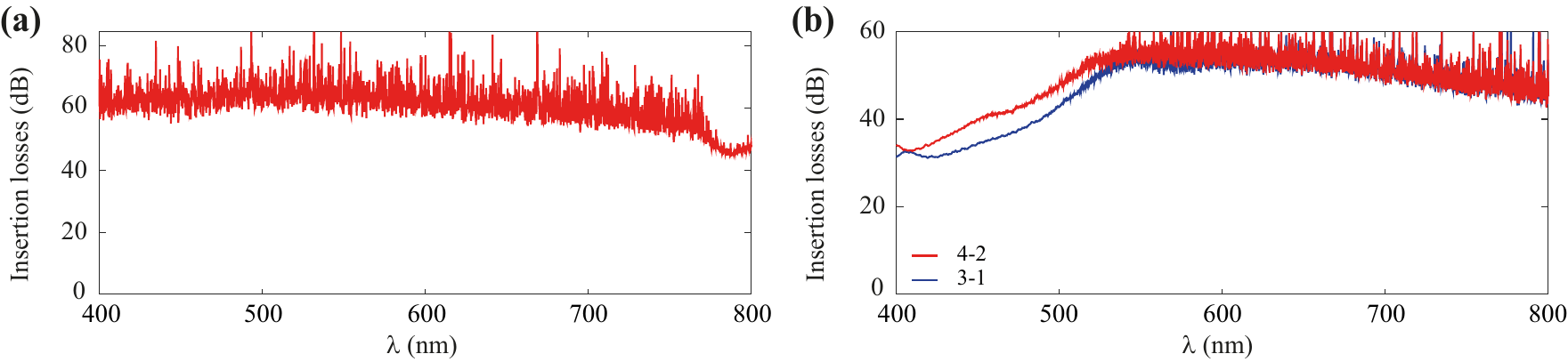}
\caption{Insertion losses measured for (a)~the dual stage isolator with light propagating from the output port to the input port; (b) the circulator (blue (red) solid lines represent insertion losses measured for light propagating from the third (fourth) port to the first (second) port).} 
\label{fig:iso}
\end{figure*}

For the isolator, the spectrum of insertion losses is shown in Fig.~\ref{fig:iso}(a). At the working (operating) wavelength (1550~nm), the measured insertion  losses for light propagating in backward direction is about 65~dB. 

From the curve depicted in Fig.~\ref{fig:iso}(a), the values of measured insertion losses are mostly below the noise level that can be detected in our experimental scheme (the dynamic range of our measurements is about $50$~dB). But, in the 772-800~nm range with the minimal losses 44.5~dB reached at $\lambda=784$~nm,  the insertion losses appear to be lower than the above value at the operating wavelength.

Similar to isolators, in circulators, losses for light propagating from one port to another (forward direction) are low, whereas the light is nearly blocked when the direction of propagation is reversed (backward propagation). In contrast to the isolators, the circulators have several ports (in our case, the circulator has four ports) that opens up additional possibilities for applications in more complicated schemes. For a number of different reasons, in real devices, the insertion losses for light propagating  in the backward direction will generally be wavelength dependent and may deviate from the value reported by manufacturers.

According to the datasheet for the circulator, at $\lambda=1550$~nm,  the insertion loss for backward direction  is about 50~dB.

Insertion losses spectra measured for the cases where light propagates from the third (fourth) port to the first (second) bypassing the second (third) port are presented in Fig.~\ref{fig:iso}(b). It can be seen that in the 400-506~nm range the losses are lower than the reported value at the working wavelength. Minimal value of insertion losses is 31.1~dB at 419~nm.

Owing to limitations of our experimental setup, we have not detected any noticeable transmission of light for backward propagation between adjacent (neighboring) ports (for instance, propagation from the second port to the first port). In this case, the results are well below the noise level and bear close resemblance to the curves measured for the isolator.


\section{IPA estimation and countermeasures}\label{sec:ipa}

\begin{figure*}[ht]
\centering
\includegraphics[width=0.9\textwidth]{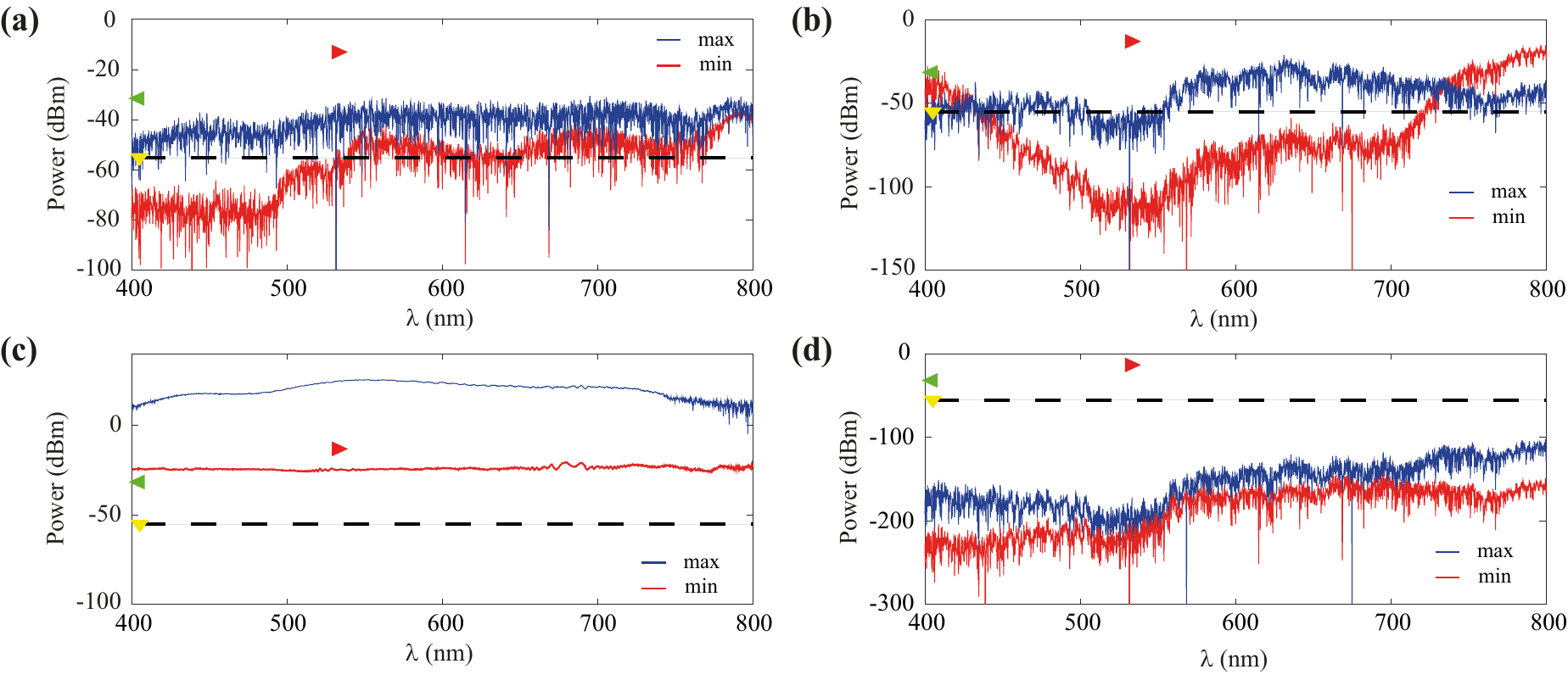}
\caption{Estimated power, Eq.~\eqref{eq:pow}, that reaches the nearest investigated modulator in a system: (a) is for the scheme from \cite{hajomer2024long},
(b) is for the scheme from \cite{lucamarini2015practical}, (c) is for the SCW QKD scheme from \cite{sajeed2021approach}, (d) is for the scheme from \cite{makarov2024preparing}; where blue (red) solid lines represents minimal (maximal) power available to an eavesdropper, black
dashed line represents conservative mean of power needed for IPA realization: yellow, red and green
triangles corresponding to estimations from \cite{ye2023induced}, \cite{han2023effect} and \cite{lu2023hacking} respectively} 
\label{fig:pow}
\end{figure*}

To show how the insertion losses influence security of QKD systems, we consider the four schemes described in Refs.~\cite{lucamarini2015practical, makarov2024preparing, hajomer2024long,sajeed2021approach} (see Fig.~5 in Ref.~\cite{lucamarini2015practical}, Fig.~1 in Ref.~\cite{makarov2024preparing}, Fig.~1 in Ref.~\cite{hajomer2024long}, and Figs~2 and~3 in~\cite{sajeed2021approach}). These schemes are ranged from the working prototype~\cite{hajomer2024long} to the ready-to-certification system~\cite{sajeed2021approach,makarov2024preparing}. They serve as examples, QKD systems that differ in the stage of development. Since comprehensive information about the models and manufacturers of all the elements used in these schemes is not accessible, we have analyzed a variety of combinations involving different elements measured in previous sections (the results of additional measurements are presented in Appendix~\ref{sec:app1}.) and have chosen the data corresponding to the minimal and maximal total insertion losses. These data are shown in Fig.~\ref{fig:pow}.

In Fig.~\ref{fig:pow}, the colored triangles indicate the powers and the corresponding wavelengths reported in Refs.~\cite{ye2023induced,han2023effect,lu2023hacking} for experimentally confirmed successful IPA. Minimal value of these powers equals 3~nW~\cite{ye2023induced}. This value can be regarded as a lower bound which is specified by the bold dashed line in Fig.~\ref{fig:pow}. We consider the cases where the power exceeds this bound as vulnerabilities. Since the IPA efficiency is known to increase when the wavelength decreases, in our analysis, vulnerabilities in the short wavelength region are treated as the most important ones.
  
Figure~\ref{fig:pow}a shows our data for the first scheme. It can be seen that, for the blue line,  
the power available to an eavesdropper is on average higher as compared to that for the red line, except for the two regions, where the red line is above the blue one.
It is important that one of these regions is the range 400-410~nm, so that the red line indicates the power enough to realize successful IPA. We consider this range as the most dangerous case. From our data for the second scheme presented in Fig.\ref{fig:pow}b we arrive at the conclusion that it is protected against IPA in all cases. Fig.~\ref{fig:pow}d shows power estimations for different optical elements in SCW QKD scheme \cite{sajeed2021approach}. According to curves shown in Fig.\ref{fig:pow}, the SCW QKD scheme is the most vulnerable to IPA system.
Note that the noise level is determined by the insertion losses of the individual elements measured, with accuracy strongly affected when the value exceeds the dynamic measurement range.
Individual outliers at certain wavelengths correspond to noise from individual pixels of the spectrometer. Therefore, the estimates can be considered conservative.

Countermeasure to this attack can be as follows. Before a connection of sender's setup to an optical
fiber link, it's voltage, applied to the modulator, should be locked by either hardware or software
(or both) and receiver's setup should calibrate only. With the proposed approach to calibration,
there is no chance that the power of prepared states will be higher than it should be -- only
lower. So again, if countermeasure is applied, IPA provides no advantage to an eavesdropper.

An alternative, straightforward countermeasure that can be implemented.
The phenomenon of IPA requires a threshold optical power to emerge.
Therefore, additional losses may be simply introduced
in order to prevent illumination of the modulator with a power higher than a threshold.

\section{Discussion and conclusion}
\label{sec:conclusion}
In this work, we investigated security vulnerabilities introduced by induced photorefraction attacks across a range of QKD protocols and proposed countermeasures where applicable. Two principal categories of concern were identified. First, we examined protocols based on linearly independent quantum states, such as phase-encoded coherent-state and SCW QKD schemes. These protocols, unlike those based on linearly dependent states, are naturally resistant to IPA-enabled phase remapping. As demonstrated in Fig.~\ref{fig:fx}, state overlap increases under phase distortion, which suppresses both unambiguous state discrimination and the success rate of any general measurement approach by an eavesdropper. The second class includes protocols that rely on amplitude or intensity modulators realized using phase modulators, which are commonly employed to control the signal power in decoy-state and SCW QKD systems. These systems remain vulnerable to IPA with partial energy recovery, emphasizing the need for careful system-level design and enhanced characterization of optical components, particularly in the visible spectral region. Thus, we have studied spectra of insertion losses of various fiber optical elements such as
optical attenuators (see Fig.~\ref{fig:Att}), WDM and isolating components (see Figs.~\ref{fig:CWDM}
and~\ref{fig:iso}, respectively) in the visible light range with wavelengths from $400$~nm to
$800$~nm. Both the optical elements and the spectral range are important for understanding
vulnerabilities of real-world QKD systems to hacking attacks, whose efficiency being wavelength
dependent may noticeably improve in the visible range.

Our experimental data show that, OAs implemented into QKD systems as countermeasures against attacks
that utilize probing pulses in the visible range require careful considerations. For VOAs, it turned
out that, at certain non-vanishing voltages, the insertion losses in the visible range are lower
than the losses at the operational wavelength (see Figs.~\ref{fig:Att}(a)-(b)). The results for the
absorption based FOAs (see Fig.~\ref{fig:Att}(c)) indicate the transmittance windows that can be
exploited by the IPA attack. By contrast, the insertion losses of the scattering based FOAs (see
Fig.~\ref{fig:Att}(d)) do not reveal noticeable changes with wavelength. So, this type of OAs is
preferable as additional countermeasure against the IPA attack.  

The measured spectra of all the WDM-components (see Fig.~\ref{fig:CWDM}) have transmittance windows
where insertion losses are not high enough to prevent the IPA. It is also shown that the insertion
losses for light propagating from the com-port to the ref-port are ranged between 10~dB and
30~dB. Since this direction is used for active monitoring purposes, such losses combined with low
efficiency of InGaAs photodiodes in the visible range will make the monitoring ineffective. One way
around this problem is to use Si-based photodiodes as the watchdog detectors against optical probing
pulses in the 400-800~nm range.  

From the insertion loss spectrum shown in Fig.~\ref{fig:iso}(a), the dual stage isolator delivers at
least 44~dB insertion losses. Presumably, such losses can be attributed to absorption of
magneto-optical material used in the isolator. Similar insertion losses were measured in the
circulator  for the case of backward propagation of light between adjacent ports. 

Insertion losses for the light passing from the third port to the first port plotted against
wavelength in Fig.~\ref{fig:iso}(b) show that the well-known broadband filtering
scheme~\cite{riant2003fiber} based on a Bragg grating connected to the second port of the circulator
becomes ineffective in the 400-506~nm range. Thus, it cannot be employed as a reliable tool to
counter the IPA. 

Countermeasures against the IPA threat fall into two primary categories.  The first is hardware-based and relatively straightforward: optical paths can be engineered to introduce additional insertion losses, thereby suppressing the intensity reaching the nearest lithium-niobate-based modulator.  Figure~\ref{fig:pow} illustrates representative estimates of the incident intensity under various configurations. In practice, combining scattering-based FOAs with a sequence of isolators can yield more than 100 dB of loss—sufficient to block probe levels required for a successful IPA. The second category is algorithmic, focusing on voltage-locking mechanisms at the transmitter’s modulators. This technique prevents unauthorized modulation adjustments, ensuring that only a trusted receiver can calibrate system parameters and  thereby reducing the overall attack surface.

\begin{acknowledgments}
B.N., A.I., A.G., A.T. and A.K. acknowledge the financial support of the Ministry of Science and Higher Education of the Russian Federation (No. FSER-2025–0007).
\end{acknowledgments}

\appendix

\section{Beamsplitters}\label{sec:app1}

\begin{figure*}[ht]
\centering
\includegraphics[width=0.9\textwidth]{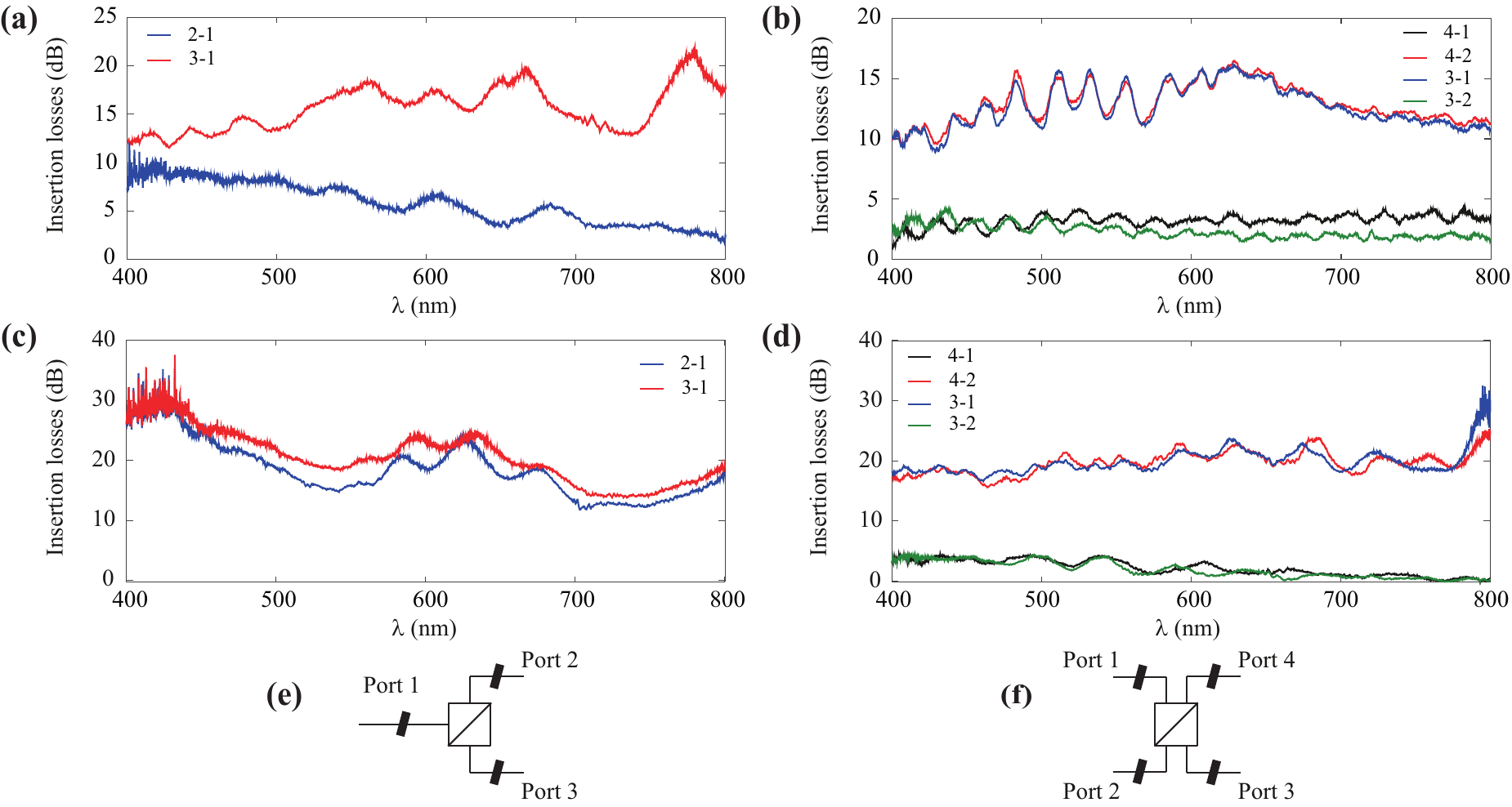}
\caption{Insertion losses measured in the 400-800 nm range for: (a) 50/50 three-port beamsplitter, (b) 50/50 four-ports beamsplitter, (c) three-port polarization beamsplitter, (d) 99/1 four-port beamsplitter, (e) and (f) illustrate ports numeration of beamsplitters} 
\label{fig:bs}
\end{figure*}

In order to present the calculated insertion losses for the investigated schemes in Fig.~\ref{fig:pow}, we additionally have measured several beamsplitters. These beamsplitters are specified for $1550$ nm wavelength: $50/50$ beamsplitter with three connection ports, Fig.~\ref{fig:bs}(a), $50/50$ beamsplitter with four connection ports, Fig.~\ref{fig:bs}(b), one polarization beamsplitter with three ports, Fig.~\ref{fig:bs}(c), and one $99/1$ with four ports, Fig.~\ref{fig:bs}(d).

To simplify the perception of the figures, only the backward propagations as the insertion losses are presented, since they have a minor difference for both the forward and the backward propagation. Here we define forward connection as the propagation of scanning light from a port with a lower number to a port with a higher number, and backward connection as its opposite, for instance, see the box in Fig.~\ref{fig:bs}.

For each beamsplitter, splitting ratios differ from the one of working wavelength, with an exception for the $99/1$ beamsplitter. This may be useful for an eavesdropper in the case when legitimate users have not measured insertion losses for beamsplitters within the considered range and have eventually chosen lower values. However, on the other hand, it may also be used as an additional countermeasure to attenuate injected light.


%

\end{document}